\documentclass[twocolumn,floatfix]{revtex4}
\pdfoutput=1
\usepackage{graphicx}
\usepackage{dcolumn}
\usepackage{longtable}
\usepackage{amsmath}
\usepackage{amssymb}
\usepackage{color}
\usepackage{isotope}

\begin{document}
\title{Islands of  shape coexistence: theoretical predictions and experimental evidence}

\author
{Andriana Martinou$^1$, Dennis Bonatsos$^1$, S.K. Peroulis$^1$, K.E. Karakatsanis$^{2,3}$, T.J. Mertzimekis$^4$, and N. Minkov$^5$ }

\affiliation
{$^1$Institute of Nuclear and Particle Physics, National Centre for Scientific Research ``Demokritos'', GR-15310 Aghia Paraskevi, Attiki, Greece}

\affiliation
{$^2$  Department of Physics, Faculty of Science, University of Zagreb, HR-10000 Zagreb, Croatia}

\affiliation
{$^3$ Physics Department, Aristotle University of Thessaloniki, Thessaloniki GR-54124, Greece}

\affiliation
{$^4$  Department of Physics, National and Kapodistrian University of Athens, Zografou Campus, GR-15784 Athens, Greece}

\affiliation
{$^5$Institute of Nuclear Research and Nuclear Energy, Bulgarian Academy of Sciences, 72 Tzarigrad Road, 1784 Sofia, Bulgaria}

\begin{abstract}

Parameter-free theoretical predictions based on a dual shell mechanism within the proxy-SU(3) symmetry of atomic nuclei, as well as  covariant density functional theory calculations  using the DDME2 functional   indicate that shape coexistence (SC) based on the particle-hole excitation mechanism cannot occur everywhere  on the nuclear chart, but is restricted on islands lying within regions of 7-8, 17-20, 34-40, 59-70, 96-112, 146-168 protons or neutrons. Systematics of data for even-even nuclei possessing $K=0$ (beta) and $K=2$ (gamma) bands support the existence of these islands, on which shape coexistence appears whenever the $K=0$ bandhead $0_2^+$ and the first excited state of the ground state band $2_1^+$ lie close in energy, with nuclei characterized by $0_2^+$ lying below the $2_1^+$ found in the center of these islands. In addition a simple theoretical mechanism leading to multiple shape coexistence is briefly discussed.    

 \end{abstract}

\maketitle
\section{Introduction}   % Sec. 1 

Shape coexistence (SC) is the term used to describe the situation in which in an  atomic nucleus the ground state band and another band lying nearby in energy possess similar energy levels, but radically different structures, for example one of them being spherical and the other one deformed, or both of them being deformed, but one of them exhibiting prolate (rugby ball like) deformation and the other one showing oblate (pancake like) deformation. First proposed in 1956 by Morinaga \cite{Morinaga}, in relation to the spectrum of $^{16}$O, SC has been observed in many odd \cite{Meyer} and even-even \cite{Wood,HW} nuclei, recently receiving intense experimental attention \cite{Garrett3} since more cases become accessible through novel radioactive ion beam facilities. 

It has been believed over the years that shape coexistence can occur practically everywhere across the nuclear chart, despite the fact that experimental examples have been appearing to cluster into islands on the nuclear chart, as seen, for example, in Fig. 8 of the review article \cite{HW}. However, it has been recently proposed, within the framework of the proxy-SU(3) symmetry \cite{proxy,proxy2}, that SC can occur only within the proton or neutron number intervals 7-8, 17-20, 34-40, 59-70, 96-112, 146-168 \cite{EPJASC,HINP2021}, which form certain horizontal and vertical stripes on the nuclear chart. Furthermore, recent covariant density functional theory (CDFT)  calculations using the DDME2 functional \cite{CDFTPLB,CDFTPRC} suggest that SC arising from nucleon particle-hole excitations occurs only within certain islands on the nuclear chart, the islands themselves lying within the stripes predicted by the proxy-SU(3) symmetry. In particular, islands of neutron-induced SC, in which SC is due to particle-hole excitations of protons, caused by the neutrons, as well as islands of proton-induced SC, in which SC is due to particle-hole excitations of neutrons, caused by the protons, have been found, in addition to islands in which both particle-hole mechanisms are present simultaneously \cite{CDFTPLB,CDFTPRC}. 

In the present work, after a brief review of the proxy-SU(3) and CDFT predictions, we consider new systematics of data which support these findings. In particular, it is shown that even-even nuclei with experimentally known $K=0$ bands \cite{ENSDF}, as well as nuclei with experimentally known $K=2$ bands  \cite{ENSDF}, form islands on the nuclear chart, within the stripes \cite{EPJASC} predicted by the proxy-SU(3) symmetry. Furthermore, it is shown that even-even nuclei in which the first excited state of the ground state band, $2_1^+$, and the first $K=0$ bandhead, $0_2^+$, lie close in energy, making these nuclei good SC candidates, also form islands on the nuclear chart, in agreement with the stripes predicted by the proxy-SU(3) symmetry. In addition, a qualitative mechanism giving rise to multiple SC \cite{Garrett4,Garrett5} is briefly described. 
  
\section{Harmonic oscillator (HO) and spin-orbit (SO) magic numbers} % Sec. 2 

Magic numbers have played a major role since the infancy of nuclear structure. 
The first set of magic numbers introduced in nuclear structure was the set of the magic numbers 2, 8, 20, 28, 50, 82, \dots, which gave rise to the shell model \cite{Mayer,Jensen}. It has been understood that these magic numbers rise from the 3-dimensional isotropic harmonic oscillator (3D-HO) magic numbers 2, 8, 20, 40, 70, 112, 168, \dots \cite{Wybourne,Smirnov,IacLie} with the addition of the spin-orbit interaction \cite{Mayer,Jensen}, which is small up to the $sd$ shell, but modifies the composition of the HO shells beyond this point.  More recent experimental evidence \cite{Sorlin} led to the distinction between the HO magic numbers 2, 8, 20, 40, 70, 112, 168, \dots, which are valid everywhere in the 
absence of any spin-orbit interaction, and spin-orbit (SO) magic numbers 6, 14, 28, 50, 82, 126, 184, \dots, \cite{Sorlin,EPJASM}, which are valid in the case of strong spin-orbit interaction everywhere.  It is clear that the shell model magic numbers are a mixture of these two sets, starting with HO magic numbers in light nuclei and ending up in medium-mass and heavy nuclei with SO magic numbers. 

\section{The dual-shell mechanism}   % sec. 3

A dual-shell mechanism \cite{EPJASC}, based on the interplay of the HO and SO shells, has been recently introduced, predicting the existence of specific regions in the nuclear chart, in which shape coexistence (SC) can occur. 

From the conceptual point of view, the dual-shell mechanism is based on the fact that the shell model magic numbers are known to be valid only at zero nuclear deformation, or close to it. When nuclear deformation comes in, the shell model magic numbers are not valid any more, as one can see in the standard Nilsson diagrams \cite{Nilsson,NR,Lederer}, in which the change of the nucleon (proton or neutron) single-particle energies as a function of the nuclear deformation is depicted. The Nilsson model \cite{Nilsson,NR} consists of a three-dimensional HO with cylindrical symmetry, to which a spin-orbit interaction is added. Despite its simplicity, it has been extremely helpful for decades in analyzing experimental data. Away from zero deformation, one can see in the Nilsson diagrams that the large gaps appearing above shell model magic numbers at zero deformation  deteriorate quickly, while new magic numbers appear and disappear with increasing deformation. The appearance of new magic numbers has recently attracted considerable attention, both from the experimental \cite{Sorlin} and theoretical \cite{Otsuka} viewpoints. 

From the algebraic point of view, the dual-shell mechanism is based on the SU(3) symmetry of the Elliott model \cite{Elliott1,Elliott2,Elliott3,Elliott4}, prevailing in light nuclei up to the $sd$ shell, and the proxy-SU(3) symmetry \cite{proxy,proxy2}, proved to be a very good approximation valid in medium mass and heavy nuclei \cite{EPJASM,Sobhani}. The dual-shell mechanism predicts that SC can be obtained only when the proton and/or the neutron numbers, $Z$ and $N$ respectively, of a given nucleus are within the intervals 7-8, 17-20, 34-40, 59-70, 96-112, 146-168. While the right borders of these regions are the well-known HO magic numbers \cite{Wybourne,Smirnov}, the left borders, 7, 17, 34, 59, 96, 146 form a new set of magic numbers (called the SC magic numbers in this work), determined by the condition that the quadrupole-quadrupole interaction within the HO and SO schemes become equal \cite{EPJASC,HINP2021}. Actually, the intervals just mentioned represent regions in which the magnitude of the quadrupole-quadrupole interaction (expressed through the second order Casimir operator of the relevant SU(3) irreducible representation (irrep) \cite{proxy2,EPJAHW,Kota}) coming from the SO shell exceeds the one coming from the HO shell. 

In more detail, in Ref. \cite{EPJASC} the third set of magic numbers (the SC magic numbers) has been determined using the eigenvalue of the second order Casimir operator of SU(3) \cite{IacLie,Kota} 
\begin{equation}
C_2(\lambda,\mu)= \lambda^2 +\mu^2 +\lambda \mu + 3 (\lambda+\mu),
\end{equation}
where $\lambda$ and $\mu$ are the Elliott quantum numbers \cite{Elliott1,Elliott2,Elliott3,Elliott4} characterizing the $(\lambda,\mu)$ irreducible representation of SU(3). The Casimir $C_2$ is known to be proportional to the quadrupole-quadrupole operator $QQ$ \cite{Elliott1,Elliott2,Elliott3,Elliott4}, as well as to the square of the collective deformation variable $\beta$ \cite{Castanos,proxy2}. For the protons (or the neutrons) of each nucleus we have two possibilities: The $(\lambda,\mu)$ irrep coming from the HO picture, or the irrep coming from the SO picture (see Table 1 of Ref. \cite{EPJASC} for ready-to-use irreps). The need to use the highest weight irreps has been emphasized in Ref. \cite{proxy2} and fully justified mathematically in Ref. \cite{EPJAHW}. The SC magic number occurs at the point at which the SO Casimir, or equivalently the QQ interaction or the $\beta$ deformation variable,  exceeds the corresponding HO one. Apparently this point is related to the onset of deformation \cite{Cakirli1,Cakirli2,Karampagia}.

If the neutron number $N$ of a given nucleus belongs to one of the above mentioned intervals, we say that this nucleus exhibits neutron-induced SC. As we shall see below, this is the kind of SC  occurring in the Pb and Hg isotopes around the $Z=82$ shell closure \cite{HW}, as well as in the Sn and Te isotopes around the $Z=50$ shell closure \cite{Wood}. In the first case, SC occurs in the interval $N=96$-112, centered around $N=104$, while in the second case, SC occurs in the interval $N=59$-70, centered around $N=64$. 

If the proton number $Z$ of a given nucleus belongs to one of the above mentioned intervals, we say that this nucleus exhibits proton-induced SC. As we shall see below, such a case occurs, for example, in the light (with $N<92$) Sm, Gd,  and Dy isotopes. SC is expected in this case just below $N=92$ and within the interval $Z=59$-72, centered around $Z=64$. 

If both the proton and the neutron numbers of a given nucleus lie within the above mentioned intervals, we say that in this nucleus SC is both proton-induced and neutron-induced. As we shall see below, this happens, for example, in the Sr and Zr isotopes in the region $Z\approx N \approx 40$. 

The regions of the nuclear chart in which SC is expected according to the dual shell mechanism are shown in Fig. 1. 
 The colored regions possess proton or neutron number in one of the the intervals $7-8$, $17-20$, $34-40$, $59-70$, $96-112$, $145-168$. The horizontal stripes correspond to the proton induced \textcolor{red}{SC}, while the vertical stripes correspond to the neutron induced \textcolor{red}{SC}.

\section{Experimental evidence for the shape coexistence (SC) magic numbers}  % sec. 4 

This third set of magic numbers is not entirely unheard of. In the authoritative review on nuclear magic numbers of Ref. \cite{Sorlin} based on experimental evidence, new gaps  corresponding to magic numbers have been identified in $N=6$, 16, 32, 34 (see \cite{Sorlin}, p. 667), which are very close to 7, 17, 34 of the third set of magic numbers mentioned above.  In the very recent review of the evolution of shell structure in exotic nuclei \cite{Otsuka}, the new magic numbers $N=16$ and $N=34$ emerged when using central and tensor forces in actual nuclei. We see therefore that the first three members of this new set of magic numbers have already been well established, from both the experimental \cite{Sorlin} and the theoretical \cite{Otsuka} viewpoint. For easy reference we shall call them the {\sl SC magic numbers}, taking advantage of the fact that they correspond to the left borders of the regions in which SC can appear \cite{EPJASC}, while the right borders are the HO magic numbers.  

Empirical evidence for the next two members (59, 96) of the SC magic numbers comes from the parabolic behavior observed when plotting certain energy levels vs. $N$.
Parabolas in plots of specific nuclear states in even nuclei, like the $2_1^+$ and $4_1^+$ states of the ground state band (gsb) \cite{Sheline,Otsuka}, or the $2_{\gamma}^+$ band-head of the $\gamma_1$ collective band \cite{Zamfir}, vs. the neutron number $N$ have been known for a long time. The parabolas are bordered by the usual shell model magic numbers. 

Parabolas also appear when plotting the $0_2^+$ states of even nuclei, usually considered as the band-heads of the relevant $\beta_1$ collective bands \cite{Wood,HW}, vs. $N$. However, in these cases the parabolas are bordered by new sets of magic numbers. $0_2^+$ states in the Sn isotopes are bordered by $N=60$ and 70 \cite{Wood}, while $0_2^+$ states in the Hg and Pb isotopes are bordered by $N=96$ and 112 \cite{HW}. In Ref. \cite{EPJASC} it has been shown that these are regions in which SC is expected to appear, bordered on the left side by SC magic numbers and on the right side by HO magic numbers, according to the terminology introduced above.   

Over the years some subshell closures, like $Z=34$, 40, 64 have received considerable attention \cite{Wood,HW}. Within the framework described above, 34 is a member of the third (SC) set of magic numbers, 40 is a member of the  set of HO magic numbers,  while 64 is the midshell of the SC candidate region 59-70.  

The empirical support for the members of the SC magic numbers discussed above,  invites into further consideration of the physics underlying them. This task will be undertaken in Sec. VI. 

\section{Islands of shape coexistence in covariant density functional theory} % sec. 5    
 
Before considering further experimental evidence in favor of the formation of islands of SC, it is worth examining what microscopic methods could say in comparison to the parameter-free symmetry-based predictions of the dual shell mechanism. In this direction, covariant density functional theory 
\cite{Ring1996,Bender2003,Vretenar2005,Meng2006,Niksic2011,Meng2015,Liang2015} calculations using the DDME2 functional \cite{Lalazissis} have been performed \cite{CDFTPLB,CDFTPRC} using the code of Ref. \cite{Niksic},   
in which a finite range paring force \cite{Tian1,Tian2} has been taken into account in its default value. The main results are shown in Fig. 2. Two islands of neutron-induced SC are seen close to the magic numbers $Z=82$ and $Z=50$, within the neutron regions in which the parabolas mentioned in Sec. IV are observed. In addition, two islands of proton-induced SC are seen close to $Z=64$ and $Z=40$ discussed in Sec. IV, while an island located near $N=Z=40$ corresponds to nuclei in which both neutron-induced and proton-induced SC are active. All islands lie entirely within the horizontal and vertical stripes predicted by the dual shell mechanism. Further discussion is deferred to Sec. IX. 

\section{Islands of $K=0$ and $K=2$ bands} % sec. 6  

A systematic collection of experimental data from the ENSDF database \cite{ENSDF} for the collective $K=0$ and $K=2$ bands in even-even nuclei throughout the nuclear chart has been performed. 

In Fig. 3 are shown the nuclei in which well-developed excited $K=0$ bands appear, also listed in Table I. All  even-even nuclei up to $Z=84$ and $N=126$ have been considered. 104 nuclei are included, fulfilling the following requirements: a) In Ref. \cite{ENSDF} these bands are denoted as $\beta$ bands, or quasi-$\beta$ bands,  or $K=0$ bands.  b) At least two levels of these bands are known experimentally.  

In Fig. 3 the regions in which shape coexistence is expected according to the dual shell mechanism, i.e., 7-8, 17-20, 34-40, 59-70, 96-112,  are also shown for both protons and neutrons. 

We see that almost all nuclei with known $K=0$ bands fall within a horizontal (proton) or vertical (neutron) zone, while the few exceptions touch  the borders of these zones.  

This fact raises several comments and questions.

1) The appearance of experimental $K=0$ bands within certain zones cannot be accidental. It is understood that searches for $K=0$ bands \cite{Garrett} have been performed over the years in other regions as well, with no $K=0$ bands been identified. 

2) $K=0$ bands are seen within proton-induced coexistence zones (horizontal zones), as well as within neutron-induced coexistence zones (vertical zones).

3) The nature of $K=0$ bands has been disputed in recent years by Sharpey-Schafer \cite{Sharpey1,Sharpey2,Sharpey} and others \cite{Garrett,Garrett2}. The present study might lead to additional insights in relation to their nature.

 In Fig. 4 are shown the nuclei in which well-developed excited $K=2$ bands appear, also listed in Table II. All  even-even nuclei up to $Z=84$ and $N=126$ have been considered. 164 nuclei are included, fulfilling the following requirements: a) In Ref. \cite{ENSDF} these bands are denoted as $\gamma$ bands, or quasi-$\gamma$ bands, or $K=2$ bands.  b) At least two levels of these bands are known experimentally. 
 
 % In 24 of these nuclei (shown by red boxes) only  levels with even $L$ are known, while  in 4 of these nuclei (shown by green boxes) only  levels with odd $L$ are known.
 
 The regions in which shape coexistence is expected are shown in the same way as in Fig. 3.
 
 We see that most of the nuclei with known $K=2$ bands fall within a horizontal zone and/or a vertical zone, as in Fig. 3. However, in the present case there are a few exceptions not touching  the borders of these zones.
 
 Comments 1) and 2) made on $K=0$ bands apply also to the $K=2$ bands. However, in contrast to the $K=0$ bands, the nature of the $K=2$ bands as being collective gamma vibrations has not been disputed \cite{stagg}.  
 
It seems therefore that almost all collective $K=0$ bands, as well as most of the $K=2$ bands,  fall within the borders of the regions proposed through the dual shell mechanism of Ref. \cite{EPJASC}. The remaining task is then to distinguish among these nuclei the ones in which SC is observed from those in which SC is not seen. In this direction, in the next section we apply the criterion that the gsb and the excited $K=0$ band should be close in energy, in order to narrow down the islands seen in Fig. 3. 

\section{Energy differences of $0_2^+$ and $2_1^+$}  % sec. 7
 
 For all nuclei exhibited in Fig. 3, in which excited $K=0$ bands known experimentally are shown,  the energy difference of the $0_2^+$ and $2_1^+$ states has been considered. The energies of the $2^+$ state of the ground state band, the $0^+$ bandhead of the $K=0$ ($\beta$) band, and the $2^+$ bandhead of the $K=2$ ($\gamma$) band, wherever known \cite{ENSDF}, are shown in Table III. Nuclei for which the energy difference $E(0_2^+) - E(2_1^+)$ remains lower than 800 keV are shown in boldface in Table I and depicted in Fig. 5 (see the last paragraph of this session for further discussion of this arbitrary choice). These are nuclei for which SC is expected to appear, since the ground state band and the  first excited $K=0$ band remain close in energy. All nuclei in which the $0_2^+$ and $2_1^+$ states come close in energy lie within the stripes predicted by the dual shell mechanism. 
 
The emerging picture is simple and clear. The nuclei in which the $0_2^+$ and $2_1^+$ states are close in energy can be divided into three classes:

a) Nuclei in which  only the neutrons belong to the intervals 7-8, 17-20, 34-40, 59-70, 96-112. These are nuclei in which  neutron-induced SC is expected.

b) Nuclei in which  only the protons belong to the intervals 7-8, 17-20, 34-40, 59-70. These are nuclei in which proton-induced SC is expected.

c) Nuclei in which both protons and neutrons belong to the intervals 7-8, 17-20, 34-40, 59-70. These are nuclei in which SC  is expected to be both proton-induced and neutron-induced. 

As an example, we assign to these classes all nuclei for which the condition $E(0_2^+) < E(2_1^+)$ is fulfilled, which appear to be the ``core nuclei'' around which islands are formed.  

a) Nuclei with neutron-induced SC: \isotope[184,186,188,194][82]{Pb}$_{102,104,106,112}$, \isotope[72][32]{Ge}$_{40}$.

b) Nuclei with proton-induced SC:  \isotope[96,98][40]{Zr}$_{56,58}$.

c) Nuclei with both neutron-induced and proton-induced SC:  \isotope[40][20]{Ca}$_{20}$. 

As a result, some nuclei which appear as border cases of SC from the one view-point, can be core cases from the other view-point.  For example, \isotope[96,98][40]{Zr}$_{56,58}$ might appear as border cases from the view-point of neutron-induced SC, since their neutron numbers lie just outside the $N=59$-70 region, but they are core cases from the view-point of proton-induced SC, since their proton number does lie within the interval $Z=34$-40. On the contrary, \isotope[72][32]{Ge}$_{40}$ might appear as a border case from the view-point of proton-induced SC, since its proton number lies just outside the $Z=34$-40 region, but it is a core case from the view-point of neutron-induced SC, since its neutron number does lie within the interval $N=34$-40.

It should be noticed that the arbitrary choice of the 800 keV cut-off for the energy difference $E(0_2^+)- E(2_1^+)$, although arbitrary, does not influence the conclusions drawn in this section. Lowering the cut-off will make the islands formed around the ``core'' nuclei having $E(0_2^+) < E(2_1^+)$ narrower, while raising the cut-off by 200 keV, for example, will make the islands formed around the ``core'' nuclei wider, but still lying within the stripes predicted by the dual shell mechanism. 

\section{A mechanism for multiple shape coexistence }  % sec. 8 

While in most cases reported in the relevant review articles \cite{Wood,HW,Garrett3}, SC of two bands is seen, multiple SC has been recently observed experimentally in $^{110,112}$Cd, in which four coexisting bands have been located \cite{Garrett4,Garrett5}. We shall see that multiple shape coexistence can occur within the dual shell mechanism mentioned earlier in a simple way. 

From the considerations made so far, it becomes plausible that in regions of moderate or large deformation, both sets of magic numbers, the SO magic numbers (6, 14, 20, 28, 50, 82, 126, \dots \cite{Sorlin}) and the 3D-HO magic numbers (2, 8, 20, 40, 70, 112, 168, \dots \cite{NR}) should play roles of comparable importance. This implies that in a given nucleus, the valence protons or neutrons could be described in SU(3) either by the relevant SO irrep or by the HO irrep.  This leads to four possible combinations for protons-neutrons: SO-SO, SO-HO, HO-SO, HO-HO. In all cases, the stretched irreps $(\lambda,\mu)=(\lambda_p+\lambda_n,\mu_p+\mu_n)$ \cite{DW1} can be used as a lowest order approximation.  Given the success of the proxy-SU(3) symmetry in predicting the $\beta$ and $\gamma$ deformation variables in many medium mass and heavy nuclei \cite{proxy2}, it is expected that the SO-SO combination would correspond to the ground  state band, while other irreps would correspond to other low-lying configurations. Since in most nuclei four very different irreps result from these combinations (or three different irreps, if equal numbers of valence protons and neutrons occupy the same shell), the structures of the corresponding bands are expected to be quite different. Therefore, if some of them lie close in energy, multiple SC would arise. It is clear that this simple mechanism can produce up to four coexisting bands. Further work is called for in this direction.

\section{Discussion}  % sec. 9   

The present work accumulates evidence that shape coexistence based on particle-hole excitations can occur only on certain islands of the nuclear chart, the main relevant points listed here.

a) Parameter-free predictions based on a dual shell mechanism within the proxy-SU(3) symmetry indicate that SC can only occur within the nucleon regions  7-8, 17-20, 34-40, 59-70, 96-112, 146-168, which form horizontal and vertical stripes on the nuclear chart. 

b) Covariant density functional theory calculations using the DDME2 functional suggest that SC based on particle-hole excitations can only occur on islands of the nuclear chart lying entirely within the proxy-SU(3) stripes mentioned above.

c) In parallel, nuclei with experimentally known collective $K=0$ and/or $K=2$ bands also lie on similar islands of the nuclear chart located mostly within the proxy-SU(3) stripes, indicating that a certain amount of collectivity is a prerequisite for the occurence of SC. 

d) Choosing from the islands of c) only the nuclei in which the $0_2^+$ and $2_1^+$ states are energetically close to each other, islands  lying entirely within the proxy-SU(3) stripes are obtained. 

In other words, the above findings indicate that SC based on particle-hole excitations can occur only on certain islands of the nuclear chart. These islands are characterized by the presence of collective $K=0$ and/or $K=2$ bands, a fact indicating that a certain minimal amount of collectivity is needed for the appearance of SC. The islands get narrower by insisting on the proximity of the energies of the $0_2^+$ and $2_1^+$ states. The islands determined in this way are in agreement with the islands on which particle-hole excitations can occur according to covariant density functional theory calculations. The islands also lie entirely within the SC stripes determined by parameter-free arguments within the proxy-SU(3) symmetry, using a dual shell mechanism.    

Looking the other way around, parameter-free arguments based on the proxy-SU(3) symmetry suggest that SC can occur only within horizontal and vertical stripes of the nuclear chart, bordered by the nucleon numbers 7-8, 17-20, 34-40, 59-70, 96-112, 146-168. Systematics of the appearance of $K=0$ and $K=2$ bands, as well as of the proximity of the $0_2^+$ and $2_1^+$ states, suggest that SC can be expected on islands lying entirely within the proxy-SU(3) stripes. This conclusion is corroborated by covariant density functional theory calculations using the DDME2 functional, indicating that particle-hole excitations can occur only within the same islands, lying within the proxy-SU(3) stripes.

The possible role of the pairing force should be mentioned at this point. Within the covariant density functional theory calculations, a finite range pairing force \cite{Tian1,Tian2} has been taken into account in its default value, while no pairing force has been included so far in the proxy-SU(3) scheme. The pairing force is known \cite{Tian1,Tian2} to improve the agreement between relativistic mean field predictions and the experimental values for the collective deformation variable $\beta$ (see Fig. 4 of Ref. \cite{Tian2}). Although the parameter-free predictions of proxy-SU(3) for $\beta$ lie in general closer to the experimental values than RMF results (see Fig. 3 of Ref. \cite{proxy2}, for example), it is clear that room for improvement through inclusion of the pairing force in the proxy-SU(3) scheme does exist.

In view of the above, the proxy-SU(3) stripes represent a necessary but not sufficient condition for the existence of SC. The stripes are narrowed down into islands by the empirically seen necessary condition of having adequate collectivity and keeping the $0_2^+$ and $2_1^+$ states close to each other. These narrower islands are corroborated by covariant density functional theory calculations using the DDME2 functional, which indicate that particle-hole excitations are found only within these islands. 

The above findings do not exclude the existence of SC in other regions, based on a different mechanism. The present results regard SC based on particle-hole excitations, for which a minimal degree of collectivity seems to be  required. 

The parameter-free proxy-SU(3) predictions  mentioned above are obtained through a dual shell mechanism, based on the collapse of the usual shell model magic numbers away from zero deformation, where two sets of magic numbers can be considered, the HO magic numbers occurring in the absence of any spin-orbit interaction, and the SO magic numbers, prevailing when the spin-orbit interaction is strong. For a given number of protons, i.e.  within a series of isotopes, the competition between the SO and HO neutron magic numbers leads to neutron-induced SC. Similarly, for a given number of neutrons, i.e. within a series of isotones, the competition between the SO and HO proton magic numbers leads to proton-induced SC. Within a more general scheme, competition between the SO and HO magic numbers can occur for both protons and neutrons, leading to multiple shape coexistence of up to four different shapes.   

\section{Conclusion}

In conclusion, both the dual shell mechanism based on the proxy-SU(3) symmetry and covariant density functional theory calculations using the DDME2 functional suggest that SC cannot occur everywhere on the nuclear chart, but is restricted within certain islands. The existence of these islands is corroborated by empirical evidence for nuclei exhibiting $K=0$ and $K=2$ bands and having closely lying $0_2^+$ and $0_1^+$ states. 

The predictions of a dual shell mechanism extended to both protons and neutrons  for the regions in which  multiple shape coexistence can occur is a project worth pursuing, in view of the recent experimental findings \cite{Garrett4,Garrett5}. 

\section*{Acknowledgements} 

Support by the Tenure Track Pilot Programme of the Croatian Science Foundation and the Ecole Polytechnique F\'{e}d\'{e}rale de Lausanne, the Project TTP-2018-07-3554 Exotic Nuclear Structure and Dynamics with funds of the Croatian-Swiss Research Programme, 
as well as by the Bulgarian National Science Fund (BNSF) under Contract No. KP-06-N48/1  is gratefully acknowledged.

%%%%%%%%%%%%%%%%%%%%%%%%%%%%%

%%%%%%%%%%%%%%%%%%%%%%%% Table 1   %%%%%%%%%%%%%%%%%%%%%%%%%%%%%%%
\begin{table*}

\caption{Nuclei with experimentally known \cite{ENSDF} well-developed $K=0$ bands, also shown in Fig. 3. For each nucleus, the angular momentum  $L$ of the highest state known in the $K=0$ band is shown. 
Based on data taken from Ref. \cite{ENSDF}. See Sec. VI for further discussion. The nuclei in which the energy difference $E(0_2^+)-E(2_1^+)$ is less than 800 keV are shown in boldface and are depicted in Fig. 5. See Sec. VII for further discussion of them. }

\bigskip

\begin{tabular}{ r r r r r r r r r r r r r r  }

\hline
nuc. & $L$ & nuc. & $L$ & nuc. & $L$ & nuc. & $L$ & nuc. & $L$ & nuc. & $L$ & nuc. & $L$ \\
 \hline
 
$^{36}$Ar     &16& \bf{$^{100}$Zr}&12& $^{124}$Xe     & 4& $^{162}$Gd     & 2& $^{166}$Yb&10& $^{176}$W      &12& \bf{$^{180}$Pt}& 6\\
\bf{$^{40}$Ar}&12& \bf{$^{100}$Mo}& 2& $^{126}$Xe     & 4& \bf{$^{154}$Dy}&12& $^{168}$Yb& 4& $^{178}$W      &22& \bf{$^{182}$Pt}& 8\\
\bf{$^{40}$Ca}& 8& \bf{$^{102}$Mo}& 2& $^{132}$Ce     & 4& \bf{$^{156}$Dy}&10& $^{170}$Yb&18& $^{182}$W      & 4& \bf{$^{184}$Pt}& 6\\
\bf{$^{42}$Ca}&12& \bf{$^{104}$Ru}& 4& $^{134}$Ce     & 2& $^{158}$Dy     & 8& $^{172}$Yb&14& $^{184}$W      & 4& \bf{$^{186}$Pt}& 6\\ 
$^{44}$Ti     & 4& \bf{$^{110}$Pd}& 6& \bf{$^{146}$Ce}& 6& $^{160}$Dy     & 4& $^{174}$Yb& 4& $^{186}$W      & 4& $^{196}$Pt     & 2\\
\bf{$^{68}$Ge}&10& \bf{$^{112}$Pd}& 4& \bf{$^{148}$Nd}& 8& $^{162}$Dy     &28& $^{176}$Yb& 2& \bf{$^{172}$Os}& 8& \bf{$^{184}$Hg}&20\\
\bf{$^{70}$Ge}& 8& \bf{$^{110}$Cd}& 6& \bf{$^{150}$Nd}& 6& $^{166}$Dy     & 2& $^{178}$Yb& 4& \bf{$^{174}$Os}& 6& \bf{$^{186}$Hg}&28\\
\bf{$^{72}$Ge}& 6& \bf{$^{112}$Cd}& 8& $^{152}$Nd     & 4& $^{156}$Er     &22& $^{168}$Hf& 4& \bf{$^{176}$Os}& 6& \bf{$^{188}$Hg}&22\\
$^{82}$Ge     & 6& \bf{$^{116}$Cd}& 2& \bf{$^{152}$Sm}&16& $^{158}$Er     & 4& $^{170}$Hf& 4& \bf{$^{178}$Os}& 6& $^{190}$Hg     & 6\\
\bf{$^{72}$Se}& 2& \bf{$^{114}$Sn}&30& $^{154}$Sm     & 6& $^{160}$Er     & 2& $^{172}$Hf& 4& \bf{$^{180}$Os}& 6& \bf{$^{184}$Pb}& 8\\
\bf{$^{76}$Kr}& 2& \bf{$^{116}$Sn}&14& \bf{$^{152}$Gd}&10& $^{162}$Er     & 2& $^{174}$Hf&26& $^{184}$Os     & 6& \bf{$^{186}$Pb}&14\\
\bf{$^{96}$Sr}&18& \bf{$^{118}$Sn}&12& \bf{$^{154}$Gd}&10& $^{164}$Er     &10& $^{176}$Hf&10& $^{186}$Os     &10& \bf{$^{188}$Pb}&14\\
\bf{$^{98}$Sr}& 6& \bf{$^{122}$Te}&18& $^{156}$Gd     &14& $^{166}$Er     &12& $^{178}$Hf& 6& $^{190}$Os     & 2& \bf{$^{194}$Pb}& 6\\
\bf{$^{96}$Zr}& 6& \bf{$^{118}$Xe}& 4& $^{158}$Gd     & 6& $^{168}$Er     & 6& $^{180}$Hf&14& $^{192}$Os     & 2& \bf{$^{196}$Pb}& 8\\
\bf{$^{98}$Zr}&12& \bf{$^{120}$Xe}& 4& $^{160}$Gd     & 4& $^{170}$Er     &22& $^{170}$W & 6& \bf{$^{178}$Pt}& 8&                &  \\

\hline

\end{tabular}
\end{table*} 

%%%%%%%%%%%%%%%%%%%%%%%% Table 2  %%%%%%%%%%%%%%%%%%%%%%%%%%%%%%%
\begin{table*}

\caption{Nuclei with experimentally known \cite{ENSDF} well-developed $K=2$ bands, also shown in Fig. 4. For each nucleus, the angular momenta  $L_e$ and $L_o$ of the highest even and odd states known in the $K=2$ band are shown. Based on data taken from Ref. \cite{ENSDF}. See Sec. VI  for further discussion. 
}

\bigskip

\begin{tabular}{ r r r r r r r r r r r r r r r r r r r r r  }

\hline
nuc. & $L_e$ & $L_o$ & nuc. & $L_e$ & $L_o$ & nuc. & $L_e$ & $L_o$ & nuc. & $L_e$ & $L_o$ & nuc. & $L_e$ & $L_o$ & nuc. & $L_e$ & $L_o$ & nuc. & $L_e$ & $L_o$ \\
 \hline
 
$^{38}$Ar& 8&  &  $^{84}$Zr& 6&21& $^{116}$Pd& 8&13& $^{134}$Ce& 8& 7& $^{158}$Dy& 8& 7& $^{166}$Hf& 2& 5& $^{188}$Os& 6& 7\\ 
$^{40}$Ca&  &13&  $^{86}$Mo& 2& 5& $^{118}$Pd& 4& 5& $^{132}$Nd& 4& 3& $^{160}$Dy&22&25& $^{168}$Hf& 6& 5& $^{190}$Os&10& 5\\
$^{58}$Fe& 4& 3&  $^{88}$Mo& 2& 5& $^{112}$Cd&10&11& $^{134}$Nd&44& 3& $^{162}$Dy&18&17& $^{170}$Hf& 4& 3& $^{192}$Os&10& 7\\
$^{62}$Zn& 6& 7& $^{102}$Zr& 4& 3& $^{110}$Te& 6&  & $^{136}$Nd& 4& 5& $^{164}$Dy&14& 9& $^{172}$Hf& 6& 5& $^{180}$Pt& 4& 9\\
$^{64}$Zn& 8&  & $^{100}$Mo& 4& 3& $^{116}$Te& 6& 5& $^{138}$Nd& 6& 5& $^{166}$Dy& 4& 5& $^{174}$Hf& 4& 5& $^{182}$Pt& 6& 7\\
$^{64}$Ge& 8&  & $^{102}$Mo& 6&  & $^{114}$Xe&20&  & $^{136}$Sm&14&  & $^{156}$Er&14&15& $^{176}$Hf& 8& 7& $^{184}$Pt& 6& 7\\
$^{66}$Ge&10&  & $^{104}$Mo&18&17& $^{116}$Xe& 6&  & $^{138}$Sm& 8& 7& $^{158}$Er&10& 7& $^{178}$Hf&14&15& $^{186}$Pt&10& 9\\
$^{68}$Ge& 4& 7& $^{106}$Mo&18&17& $^{118}$Xe&14&11& $^{138}$Gd&22&  & $^{160}$Er& 2&13& $^{180}$Hf&20&15& $^{188}$Pt&24& 3\\
$^{70}$Ge& 8& 5& $^{108}$Mo&12&11& $^{120}$Xe&28& 5& $^{140}$Gd& 8& 7& $^{162}$Er&12&11& $^{170}$W & 4& 3& $^{190}$Pt& 6& 5\\
$^{68}$Se&14&  & $^{110}$Mo& 4& 5& $^{122}$Xe&14&15& $^{140}$Xe&  &13& $^{164}$Er&18&21& $^{176}$W & 4& 5& $^{192}$Pt& 8& 7\\
$^{70}$Se&14&  & $^{100}$Ru& 8& 9& $^{124}$Xe&10&21& $^{146}$Ce& 4& 5& $^{166}$Er&14&13& $^{178}$W & 4& 5& $^{194}$Pt& 4& 5\\
$^{74}$Se& 6&19& $^{102}$Ru&10&11& $^{126}$Xe& 8& 7& $^{148}$Nd& 6& 3& $^{168}$Er&12& 7& $^{180}$W & 6& 7& $^{196}$Pt& 8& 5\\
$^{76}$Se& 8& 9& $^{104}$Ru& 8& 7& $^{128}$Xe& 6& 5& $^{150}$Nd& 4& 3& $^{170}$Er&18&19& $^{182}$W & 8& 5& $^{198}$Pt& 8&  \\
$^{78}$Se&10& 9& $^{106}$Ru& 8& 7& $^{122}$Ba& 4&11& $^{152}$Sm&12&11& $^{172}$Er& 2& 5& $^{184}$W &12& 5& $^{182}$Hg&20&  \\
$^{74}$Kr&  & 7& $^{108}$Ru&10&13& $^{124}$Ba&10&11& $^{154}$Sm& 6& 7& $^{160}$Yb& 2& 3& $^{186}$W &12& 5& $^{186}$Hg& 2& 7\\
$^{76}$Kr& 8& 9& $^{110}$Ru&14&15& $^{126}$Ba&16& 9& $^{152}$Gd& 8&13& $^{162}$Yb& 2& 3& $^{188}$W & 4&  & $^{180}$Pb& 8&  \\
$^{78}$Kr&10& 9& $^{112}$Ru&16&19& $^{128}$Ba&14& 9& $^{154}$Gd& 6& 7& $^{164}$Yb& 4&13& $^{172}$Os& 6& 5& $^{182}$Pb&12&  \\
$^{80}$Kr& 6& 7& $^{114}$Ru& 6& 9& $^{130}$Ba&10& 5& $^{156}$Gd&12&15& $^{166}$Yb&12&13& $^{174}$Os& 4& 5& $^{190}$Pb& 8&  \\
$^{82}$Kr& 4& 5& $^{116}$Ru& 6& 5& $^{132}$Ba&12& 7& $^{158}$Gd& 6& 5& $^{168}$Yb& 6&29& $^{176}$Os& 4& 5& $^{194}$Po& 6&  \\
$^{84}$Kr&10&  & $^{118}$Ru& 4& 3& $^{124}$Ce&32&  & $^{160}$Gd&12& 7& $^{170}$Yb&16&17& $^{178}$Os& 4& 5& $^{198}$Po& 4&  \\
$^{80}$Sr&10&21& $^{108}$Pd&10& 7& $^{126}$Ce& 4& 3& $^{162}$Gd& 4& 3& $^{172}$Yb& 4& 5& $^{180}$Os& 6& 9&           &  &  \\
$^{82}$Sr&20& 9& $^{110}$Pd& 8& 5& $^{128}$Ce&18& 9& $^{166}$Gd& 4& 5& $^{174}$Yb& 4& 5& $^{182}$Os& 6& 7&           &  &  \\
$^{84}$Sr& 4& 9& $^{112}$Pd&10&13& $^{130}$Ce&26&  & $^{154}$Dy& 6& 7& $^{176}$Yb& 4& 5& $^{184}$Os& 6& 5&           &  &  \\
$^{82}$Zr&  &13& $^{114}$Pd&10&13& $^{132}$Ce&14& 5& $^{156}$Dy&12&15& $^{156}$Hf& 6&  & $^{186}$Os&12&13&           &  &  \\

\hline

\end{tabular}
\end{table*} 

%%%%%%%%%%%%%%%%%%%%%%%% Table 3   %%%%%%%%%%%%%%%%%%%%%%%%%%%%%%%
\begin{table*}

\caption{Nuclei with experimentally known \cite{ENSDF} well-developed $K=0$ bands, also shown in Table I and Fig. 3. For each nucleus, the energies of the $2^+$ state of the ground state band ($2_g^+$), the $0^+$ state of the $K=0$ ($\beta$) band ($0_\beta^+$), and the $2^+$ state of the $K=2$ ($\gamma$) band ($2_\gamma^+$), taken from Ref. \cite{ENSDF},  are shown in keV. 
The nuclei in which the energy difference $E(0_2^+)-E(2_1^+)$ is less than 800 keV are shown in boldface and are depicted in Fig. 5. See Sec. VII for further discussion.}

\bigskip

\begin{tabular}{ r r r r r r r r r r r r r r r r  }

\hline
nuc. & $2_g^+$ & $0_\beta^+$& $2_\gamma^+$ & nuc. & $2_g^+$ & $0_\beta^+$& $2_\gamma^+$ & nuc. & $2_g^+$ & $0_\beta^+$& $2_\gamma^+$ & nuc. & $2_g^+$ & $0_\beta^+$& $2_\gamma^+$ \\
 \hline

$^{36}$Ar      &1970.4&4329.1&      & \bf{$^{118}$Sn}&1229.7&1758.3&      & $^{156}$Er& 344.5& 930.1& 930.5& $^{184}$W      & 111.2&1002.5& 903.3\\                   
\bf{$^{40}$Ar} &1460.8&2120.9&      & \bf{$^{122}$Te}& 564.1&1357.4&      & $^{158}$Er& 192.2& 806.4& 820.1& $^{186}$W      & 122.6& 883.6& 738.0\\               
\bf{$^{40}$Ca} &5248.8&3352.6&      & \bf{$^{118}$Xe}& 337.3& 830.4& 928.1& $^{160}$Er& 125.8& 893.6& 854.4& \bf{$^{172}$Os}& 227.8& 758.3& 918.8\\                
\bf{$^{42}$Ca} &1524.7&1837.3&      & \bf{$^{120}$Xe}& 322.6& 908.7& 876.1& $^{162}$Er& 102.0&1087.2& 900.7& \bf{$^{174}$Os}& 158.6& 545.3& 846.2\\              
$^{44}$Ti      &1083.1&1904.3&      & $^{124}$Xe     & 354.0&1268.9& 846.5& $^{164}$Er&  91.4&1246.0& 860.3& \bf{$^{176}$Os}& 135.1& 601.2& 863.6\\                                 
\bf{$^{68}$Ge} &1015.8&1754.5&1777.4& $^{126}$Xe     & 388.6&1313.9& 879.9& $^{166}$Er&  80.6&1460.0& 785.9& \bf{$^{178}$Os}& 132.2& 650.5& 864.4\\                                   
\bf{$^{70}$Ge} &1039.5&1215.6&1707.7& $^{132}$Ce     & 325.3&1158.4& 822.2& $^{168}$Er&  79.8&1217.2& 821.2& \bf{$^{180}$Os}& 132.1& 736.4& 870.4\\                                    
\bf{$^{72}$Ge} & 834.0& 691.4&      & $^{134}$Ce     & 409.2&1533.5& 965.7& $^{170}$Er&  78.6& 890.9& 934.0& $^{184}$Os     & 119.8&1042.0& 942.9\\                                  
$^{82}$Ge      &1348.3&2333.6&      & \bf{$^{146}$Ce}& 258.5&1043.2&1381.9& $^{166}$Yb& 102.4&1043.0& 932.4& $^{186}$Os     & 137.2&1061.0& 767.5\\                                    
\bf{$^{72}$Se} & 862.1& 937.2&      & \bf{$^{148}$Nd}& 301.7& 916.9&1248.9& $^{168}$Yb&  87.7&1155.2& 984.0& $^{190}$Os     & 186.7& 911.8& 558.0\\                                   
\bf{$^{76}$Kr} & 424.0& 769.9&1221.7& \bf{$^{150}$Nd}& 130.2& 675.9&1062.1& $^{170}$Yb&  84.3&1069.4&1145.7& $^{192}$Os     & 205.8& 956.5& 489.1\\                                     
\bf{$^{96}$Sr} & 814.9&1229.3&      & $^{152}$Nd     &  72.4&1139.0&      & $^{172}$Yb&  78.7&1042.9&1465.9& \bf{$^{178}$Pt}& 170.3& 421.0&      \\                                  
\bf{$^{98}$Sr} & 144.2& 215.6&      & \bf{$^{152}$Sm}& 121.8& 684.8&1085.8& $^{174}$Yb&  76.5&1487.1&1634.0& \bf{$^{180}$Pt}& 153.2& 478.1& 677.5\\                                
\bf{$^{96}$Zr} &1750.5&1581.6&      & $^{154}$Sm     &  82.0&1099.3&1440.0& $^{176}$Yb&  82.1&1139.0&1260.9& \bf{$^{182}$Pt}& 155.0& 499.7& 667.8\\                                    
\bf{$^{98}$Zr} &1222.9& 854.0&      & \bf{$^{152}$Gd}& 344.3& 615.4&1109.2& $^{178}$Yb&  84.0&1315.0&      & \bf{$^{184}$Pt}& 163.0& 491.8& 648.7\\                                                  
\bf{$^{100}$Zr}& 212.5& 331.1&      & \bf{$^{154}$Gd}& 123.1& 680.7& 996.3& $^{168}$Hf& 124.1& 942.1& 875.9& \bf{$^{186}$Pt}& 191.5& 471.5& 607.2\\
\bf{$^{100}$Mo}& 535.6& 695.1&1063.8& $^{156}$Gd     &  89.0&1049.5&1154.2& $^{170}$Hf& 100.8& 879.6& 961.3& $^{196}$Pt     & 355.7&1135.3& 688.7\\
\bf{$^{102}$Mo}& 296.6& 698.3& 847.9& $^{158}$Gd     &  79.5&1196.2&1187.1& $^{172}$Hf&  95.2& 871.3&1075.3& \bf{$^{184}$Hg}& 366.8& 375.1&      \\
\bf{$^{104}$Ru}& 358.0& 988.3& 893.1& $^{160}$Gd     &  75.3&1325.7& 988.4& $^{174}$Hf&  91.0& 828.1&1226.8& \bf{$^{186}$Hg}& 405.3& 523.0&1096.6\\
\bf{$^{110}$Pd}& 373.8& 946.7& 931.2& $^{162}$Gd     &  71.6&1427.0& 864.0& $^{176}$Hf&  88.3&1149.9&1341.3& \bf{$^{188}$Hg}& 412.8& 824.5&      \\
\bf{$^{112}$Pd}& 348.7&1125.5& 813.6& \bf{$^{154}$Dy}& 334.3& 660.6&1027.0& $^{178}$Hf&  93.2&1199.4&1174.6& $^{190}$Hg     & 416.3&1278.6&      \\
\bf{$^{110}$Cd}& 657.8&1473.1&      & \bf{$^{156}$Dy}& 137.8& 675.6& 890.5& $^{180}$Hf&  93.3&1101.9&1199.7& \bf{$^{184}$Pb}&      & 570.0&      \\
\bf{$^{112}$Cd}& 617.5&1224.3&1312.4& $^{158}$Dy     &  98.9& 990.5& 946.3& $^{170}$W & 156.7&      & 937.1& \bf{$^{186}$Pb}&      & 655.0&      \\
\bf{$^{116}$Cd}& 513.5&1282.6&      & $^{160}$Dy     &  86.8&1279.9& 966.2& $^{176}$W & 108.3& 843.3&1040.2& \bf{$^{188}$Pb}& 723.6& 591.0&      \\
\bf{$^{114}$Sn}&1299.9&1953.3&      & $^{162}$Dy     &  80.7&1400.3& 888.2& $^{178}$W & 105.9& 997.0&1110.4& \bf{$^{194}$Pb}& 965.1& 930.7&      \\
\bf{$^{116}$Sn}&1293.6&1756.9&      & $^{166}$Dy     &  76.6&1149.0& 857.2& $^{182}$W & 100.1&1135.8&1221.4& \bf{$^{196}$Pb}&1049.2&1142.9&      \\

\hline

\end{tabular}
\end{table*}

%%%%%%%%%%%%%%%%%%%%%%%%%%%%%%%%%%%%%%%%%%% FIG. 1 %%%%%%%%%%%%%%%%%%%%%%%%%%%%%%%%%%%%%%%%%%%%%

\begin{turnpage}
\begin{figure} 
    \centering
    \includegraphics[scale=0.2]{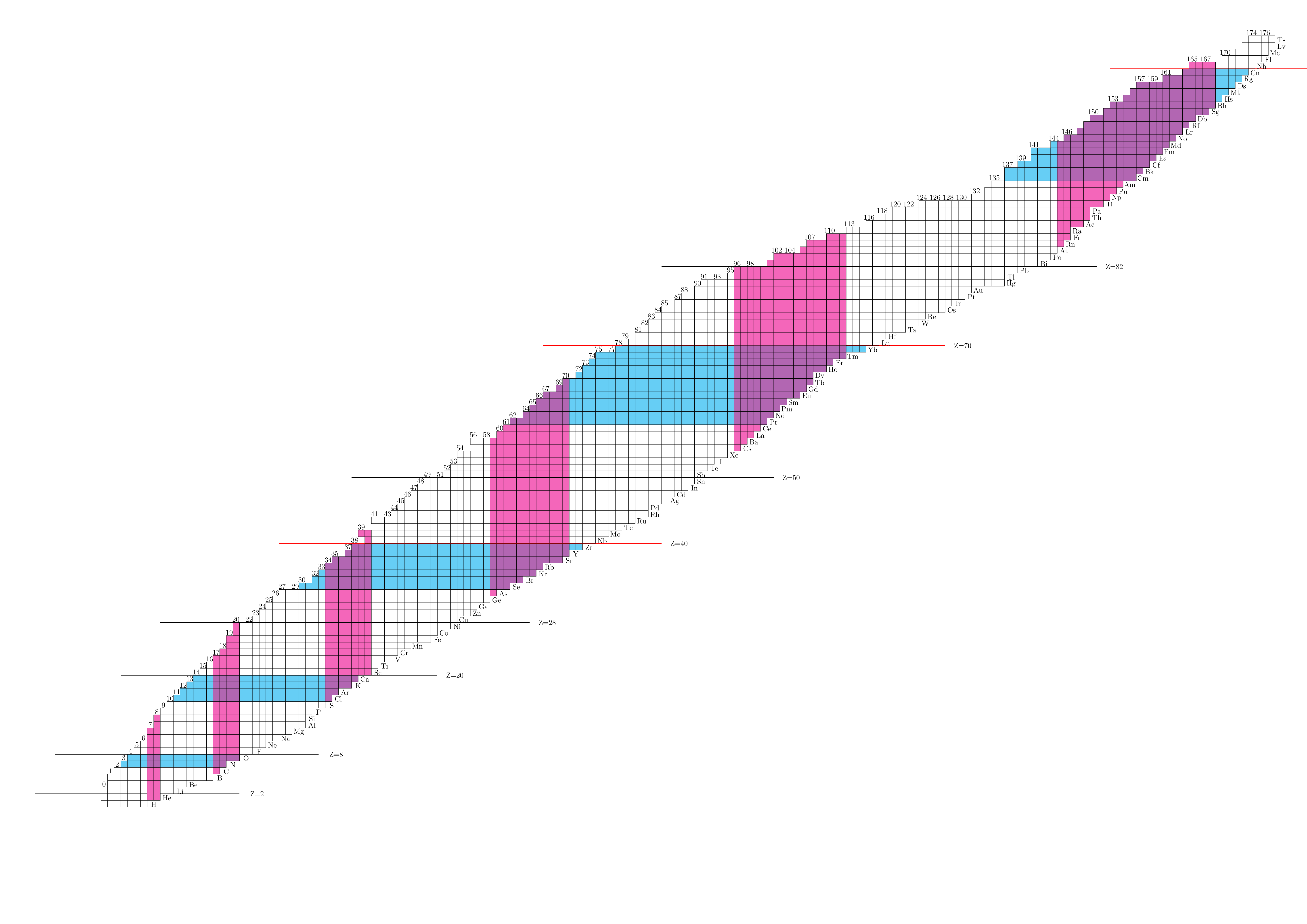}
    \caption{ Regions in which shape coexistence can appear, according to the dual shell mechanism of Sec. III. The colored regions possess proton or neutron number within the intervals $7-8$, $17-20$, $34-40$, $59-70$, $96-112$, $145-168$. The horizontal stripes correspond to the proton induced SC, while the vertical stripes correspond to the neutron induced SC. Adapted from Ref. \cite{EPJASC}. See Sec. III for further discussion.}
    \label{map}
\end{figure}
\end{turnpage}

%%%%%%%%%%%%%%%%%%%%%%%%%%% FIG. 2 %%%%%%%%%%%%%%%%%%%%%%%%%%%%%%%%%%%%%%%%%%%

\begin{figure*}[htb]

\includegraphics[width=150mm]{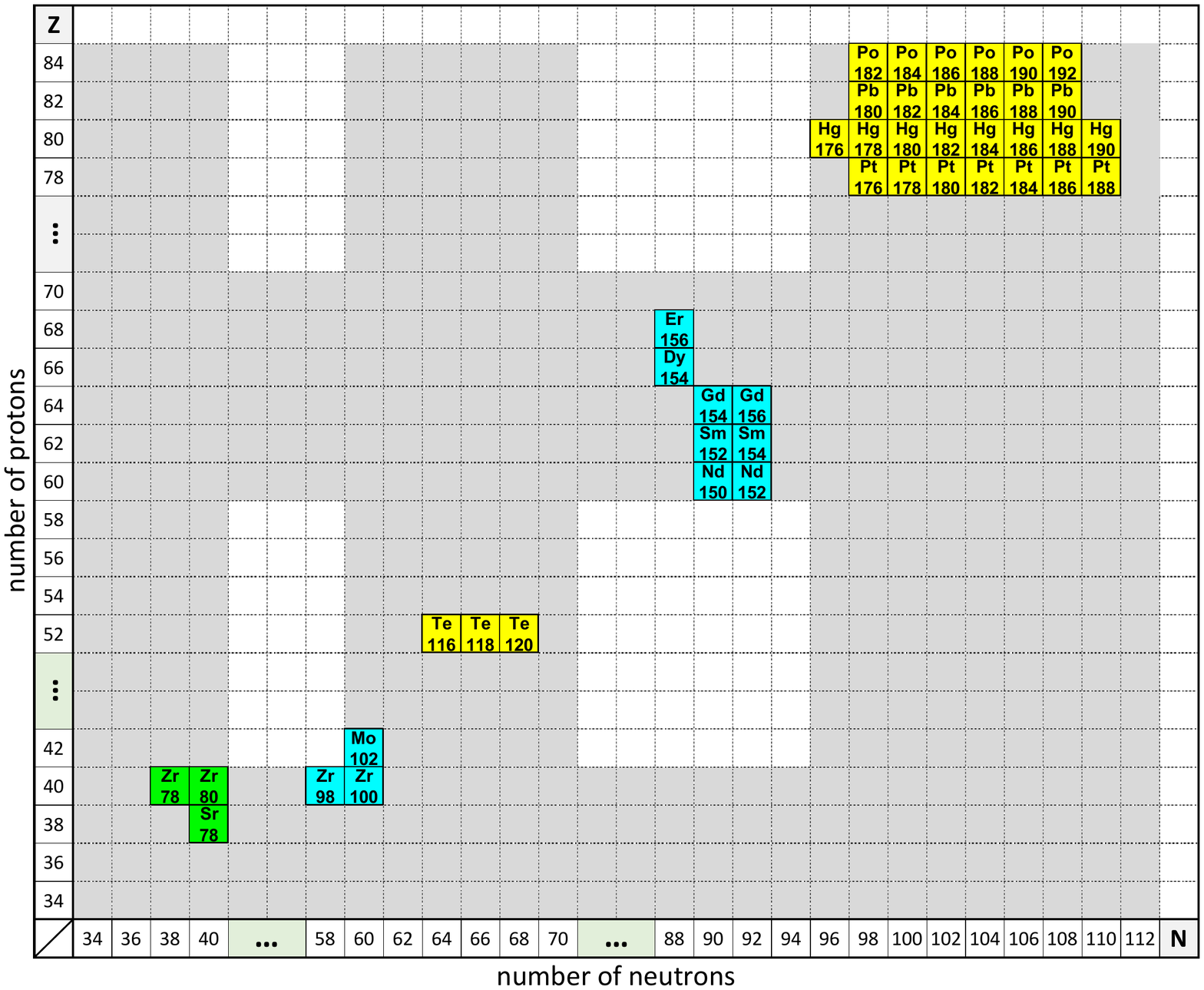}

\caption{Islands of shape coexistence found through covariant density functional theory calculations, as described in Sec. V. Islands corresponding to neutron-induced SC are shown in yellow, islands due to proton-induced SC are exhibited in azure, while islands due to both mechanisms are shown in green. The stripes in which SC is allowed according to the dual shell mechanism are shown in grey.  Adapted from Ref. \cite{CDFTPRC}. See Sec. V for further discussion. }

\end{figure*}

%%%%%%%%%%%%%%%%%%%%%%%%%%% FIG. 3 %%%%%%%%%%%%%%%%%%%%%%%%%%%%%%%%%%%%%%%%%%%

\begin{figure*}[htb]

\includegraphics[width=180mm]{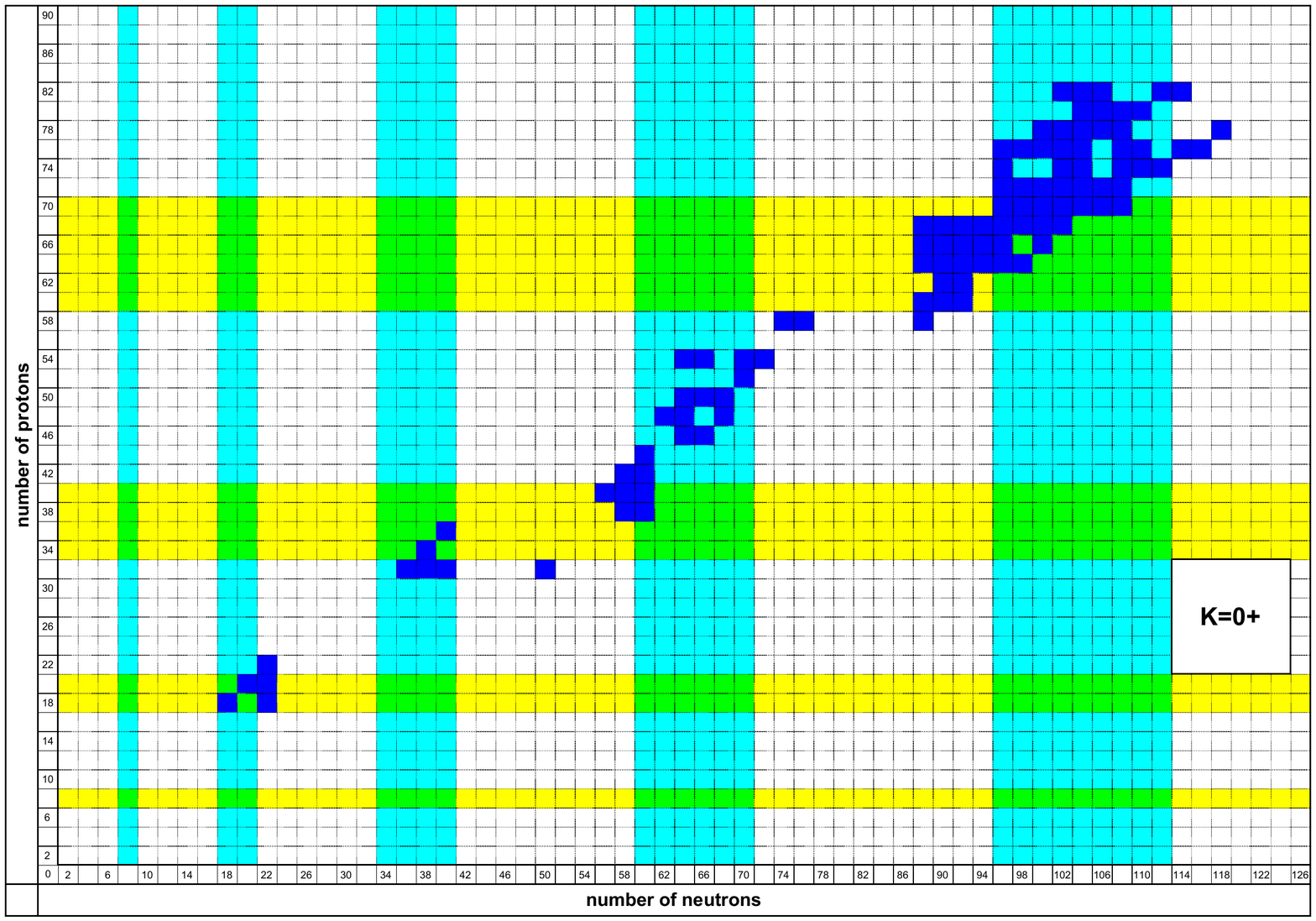}

\caption{Nuclei with well-developed $K=0$ bands (blue boxes), also listed in Table I, are given on the proton-neutron map including the stripes within which proton-induced (yellow) or neutron-induced (azure) SC can be expected. Based on data taken from Ref. \cite{ENSDF}.  See Sec. VI for further discussion. 
}  

\end{figure*}

%%%%%%%%%%%%%%%%%%%%%%%%%%% FIG. 4  %%%%%%%%%%%%%%%%%%%%%%%%%%%%%%%%%%%%%%%%%%%

\begin{figure*}[htb]  

\includegraphics[width=180mm]{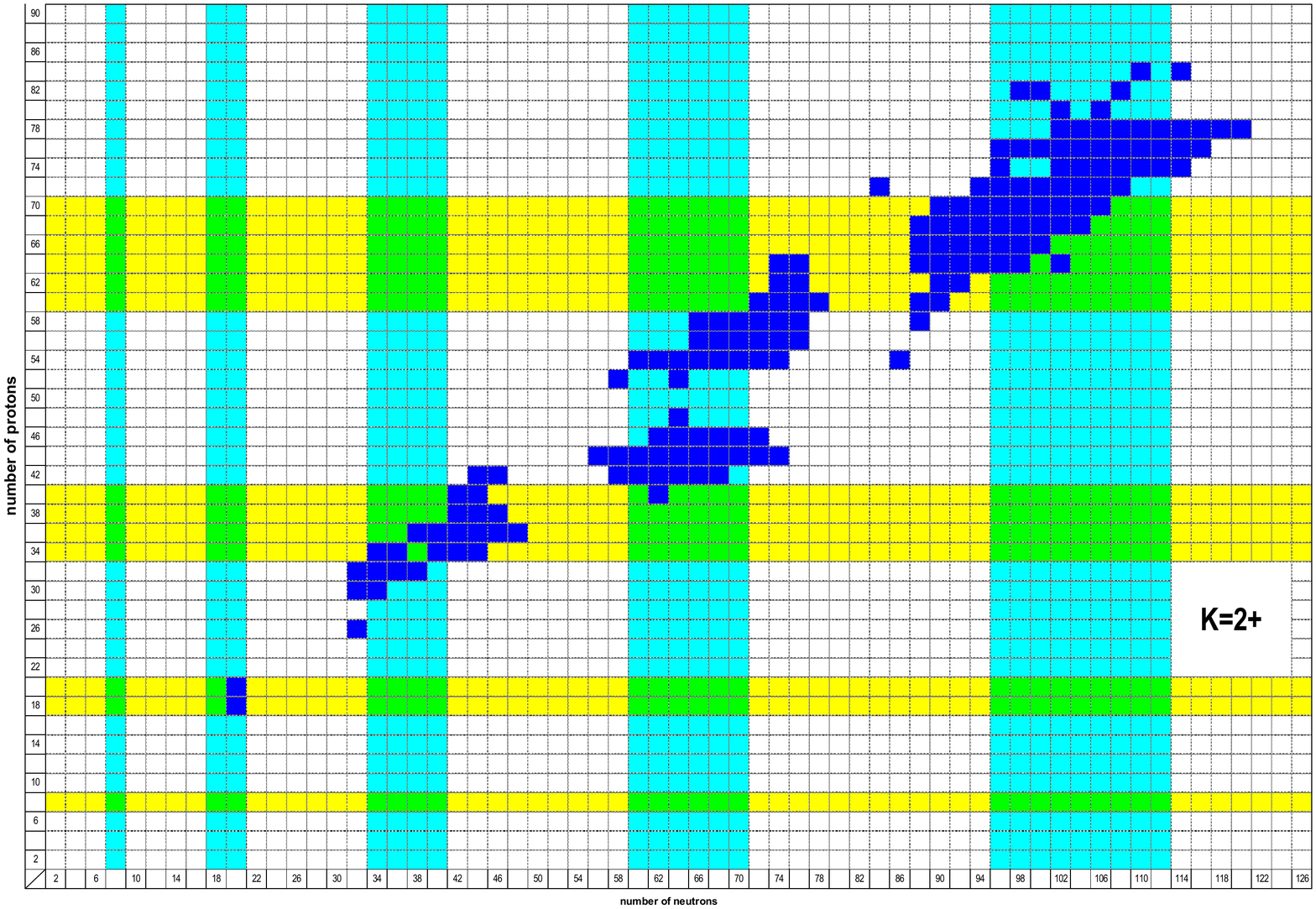}

\caption{Nuclei with well-developed $K=2$ bands (blue boxes), also listed in Table II, are given on the proton-neutron map including the stripes within which proton-induced (yellow) or neutron-induced (azure) SC can be expected. Based on data taken from Ref. \cite{ENSDF}. See Sec. VI for further discussion. 
} 

\end{figure*}

%%%%%%%%%%%%%%%%%%%%%%%%%%% FIG. 5  %%%%%%%%%%%%%%%%%%%%%%%%%%%%%%%%%%%%%%%%%%%

\begin{figure*}[htb]  

\includegraphics[width=180mm]{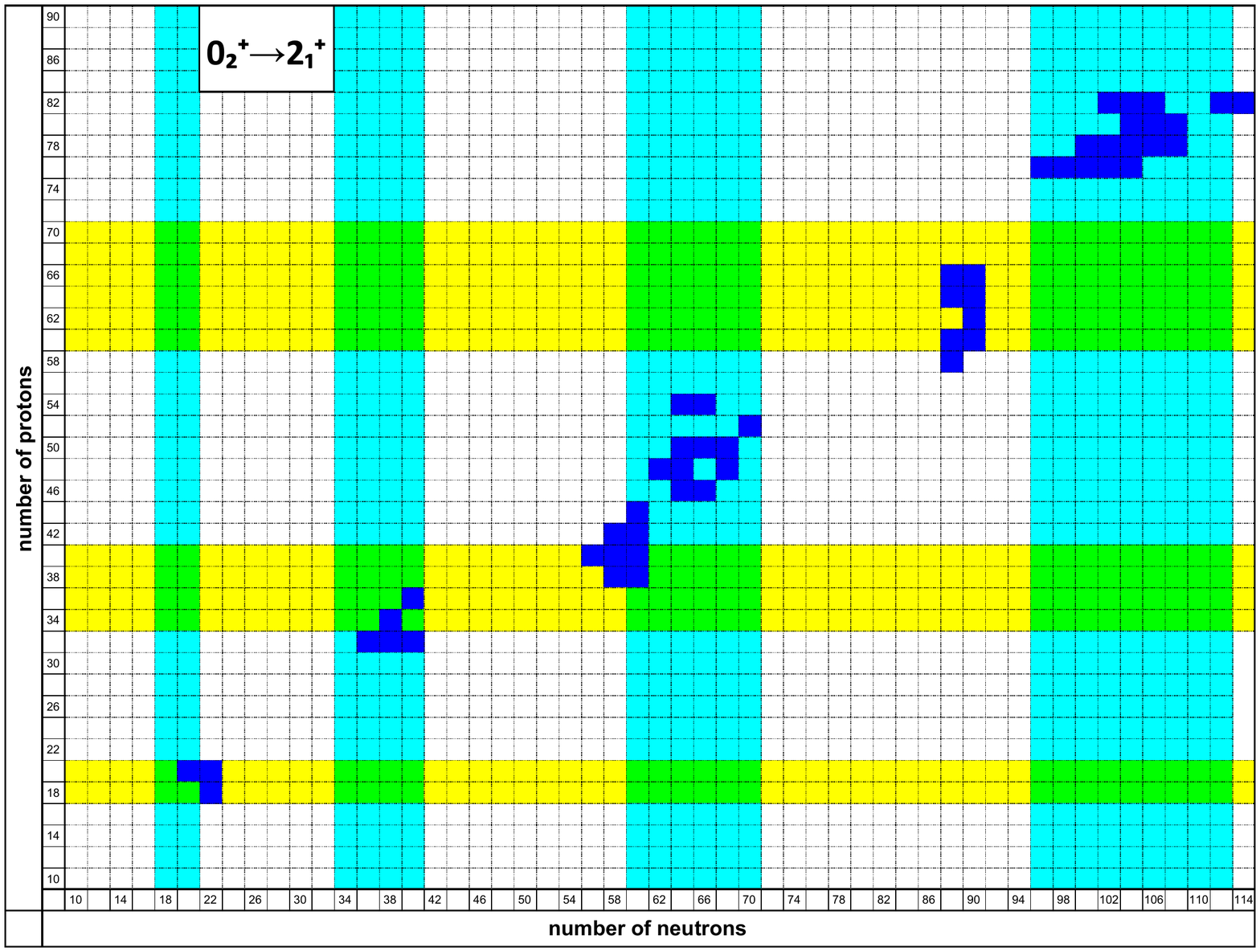}

\caption{Nuclei with energy differences $E(0_2^+)-E(2_1^+)$ less that 800 keV (blue boxes), are given on the proton-neutron map including the stripes within which proton-induced (yellow) or neutron-induced (azure) SC can be expected. Based on data taken from Ref. \cite{ENSDF}. See Sec. VII for further discussion. 
} 

\end{figure*}

\end{document}